# Determination of volume, shape and refractive index of individual blood platelets

I.V. Kolesnikova[a,b], S.V. Potapov[b], M. A. Yurkin[b,c], A.G. Hoekstra[c], V.P. Maltsev[a,b] and K.A. Semyanov[b,*]

[a]*Novosibirsk State University, Pirogova Str. 2, 630090, Novosibirsk, Russia*
[b]*Institute of Chemical Kinetics and Combustion, Siberian Branch of the Russian Academy of Sciences, Institutskaya Str. 3, 630090, Novosibirsk, Russia*
[c]*Faculty of Science, Section Computational Science, of the University of Amsterdam, Kruislaan 403, 1098 SJ, Amsterdam, The Netherlands*

**Abstract**

Light scattering patterns (LSP) of blood platelets were theoretically and experimentally analyzed. We used spicular spheroids as a model for the platelets with pseudopodia. The Discrete Dipole Approximation was employed to simulate light scattering from an individual spicular spheroid constructed from a homogeneous oblate spheroid and 14 rectilinear parallelepipeds rising from the cell centre. These parallelepipeds have a weak effect on the LSP over the measured angular range. Therefore a homogeneous oblate spheroid was taken as a simplified optical model for platelets. Using the T-matrix method, we computed the LSP over a range of volumes, aspect ratios and refractive indices. Measured LSPs of individual platelets were compared one by one with the theoretical set and the best fit was taken to characterize the measured platelets, resulting in distributions of volume, aspect ratio and refractive index.

[*] Corresponding author. Tel.: +7 383 3333240; fax: +7 383 3307350.
 *e-mail address*: kostik@kinetics.nsc.ru

## 1. Introduction

Platelets are a small subcellular fragment of blood, and have an important role in maintaining haemostatic and arterial thrombosis, particularly in fibrillation. Activated platelets adhere to an injury and aggregate to form a haemostatic plug or thrombus. Characteristics such as volume, size and shape are markers of platelet activation and function and can be used to e.g. distinguish activated - from non-activated platelets. That is why human platelets are frequently used as a tool for investigation of disease, such as ischaemic heart disease, cerebrovascular disease, renovascular disease and etc. [1]

Up to now a relation between the volume of platelets of normal healthy volunteers versus those of patients has not been revealed. However, increased volume and reduced count have been observed for 60 percent of patients who suffer from heart ishaemic disease [1]. People who had an elevated Mean Platelet Volume (MPV) at 6 months post-infarction were likely to die or suffer a further ishaemic heart event over the subsequent 18 months. MPV (standard deviation) in a control group and patients with vascular disease amounted to 8.3 (1.0) femtoliters (fl) and 9.1 (0.8) fl associated with acute myocardial infarction, 8.9 (0.9) fl and 11.3 (1.3) fl associated with ishaemic stroke, 7.2 (0.8) fl and 8.0 (1.5) fl associated with diabetes mellitus. From these numbers one can conclude that platelet volume alone is not sufficient to unambiguously detect the type of disease.

Unfortunately platelet volume is currently the only morphological characteristic routinely measured in the majority of hematology laboratories. Laboratories use volume measurement systems that utilize different physical principles. For example, some systems use aperture-impedance techniques (e.g. Coulter S Plus) [2], while others use optical techniques (e.g. Technicon H6000) [3, 4]. Both systems do not take into account the shape variability of platelets. As a result, both techniques do not agree with each other [5]. Moreover, different anticoagulants added to the solution can influence the shape of platelets in different ways [6]. In this study we propose a method to estimate the volume *and* shape of platelets under different conditions. We believe that this enhanced characterization of platelets may have clinical relevance.

A Scanning Flow Cytometer (SFC) [7, 8] allows measurement of the angular dependence of light-scattering intensity (the light-scattering pattern, LSP) of individual particles. The LSP of a platelet contains information about its morphology parameters (internal structure, shape and size) as well as its refractive index. In order to extract information from a platelet's LSP it is necessary to solve the inverse light-scattering problem. Unfortunately, methods to solve it for particles more complex than a sphere hardly exist. The only existing approach to determine characteristics of complex particles from LSP is the direct comparison of experimentally measured LSP and theoretically calculated LSPs based on its optical model. We have demonstrated the validity of this approach for optical characterization of red blood cells [9].

In EDTA solution platelets are spicular spheroids [6]. In order to evaluate the influence of the pseudopodia on the light scattering we used the Discrete Dipole Approximation (DDA) to simulate LSP of a spicular spheroid.

Theoretical calculations were carried out with the T-matrix method for oblate spheroids for a range of volumes, aspect ratios and refractive indices. Theoretical and experimental LSPs of platelets were compared one by one, and $\chi^2$ values were calculated. The characteristics of a measured platelet were determined from the one spheroid whose theoretical LSP gave minimal $\chi^2$ difference. Distributions of volume, aspect ratio and refractive index of platelets were produced.

## 2. Platelet sample preparation

Blood was withdrawn from the antecubital vein of normal volunteer with a syringe and placed into a tube filled with EDTA as an anticoagulant at a ratio of 1.5 mg EDTA to 1 ml blood in sample. All surfaces in contact with the blood were plastic. The sample was stored at room temperature. The measurement was conducted in about 30 minutes after venesection. The same sample was measured with commercial cell counter (Coulter MAX M, Beckman Coulter Corporation).

## 3. Optic model of platelets

In the activated state a platelet is a spicular spheroid and in the nonactivated state a discoid cell with diameter 2 - 4 μm and thickness 0.5 - 2 μm. Activation of platelets starts when oxygen appears in the blood plasma. The process of platelet activation includes transformation of the cell to aggregate into a thrombus – pseudopodia emerge from a discoid platelet which becomes spicular [10].

The problem of platelet measurements relates to variation of its volume, associated with anticoagulants such as EDTA, which cause the volume to elevate, the inner structure to change and platelets themselves to become spheroid [6]. The reason probably is the visible swelling of the platelet canalicular system. Platelets in sodium–citrate showed minimal changes in structure, almost without activation. In EDTA, a typical swelling of the canalicular system was registered. Platelet shape changes are governed by changes in the bilayer balance. Whenever platelets are treated with agents that expand the plasma membrane outer monolayer, relative to the inner monolayer, platelets become spicular and the canalicular system less prominent. Expansion of the inner monolayer causes a loss of spicules and a larger prominence of the canalicular system [11].

Although platelets in EDTA solution are inhomogeneous spicular spheroids, we propose to model them as homogeneous spheroids. Our approximation does not take into account pseudopodia on the surface and organelles inside the platelet, but it allows us to estimate a platelet volume, shape (aspect ratio) and optical density of the internal substance. In order to evaluate the influence of pseudopodia on light scattering we have constructed an optical model of a homogeneous oblate spheroid with 14 rectilinear parallelepipeds. The parallelepipeds are of square section, straight and coming from the center in the direction of the face centers and corners of a circumscribing cube (Fig. 1).

## 4. DDA simulation

The basics of the Discrete Dipole Approximation (DDA) were summarized by Draine and Flatau [12]. Our code – Amsterdam DDA (ADDA) – is capable of running on a cluster of computers (parallelizing a single DDA computation), which allows us to use practically an unlimited number of dipoles, since we are not limited by the memory of a single computer [9, 13]. In this paper we use the Lattice Dispersion Relation prescription for dipole polarizability [14] and Quasi Minimal Residual method [15] for iterative solving of the linear system.

We used 12.7 dipoles per wavelength, a 128×128×128 computational box with 2097152 dipoles. The LSPs were simulated with a typical time of 25 seconds on a cluster of 32 P4-3.4 GHz processors. All DDA simulations were carried out on the Netherlands national compute cluster LISA[1].

---
[1] http://www.sara.nl/userinfo/lisa/

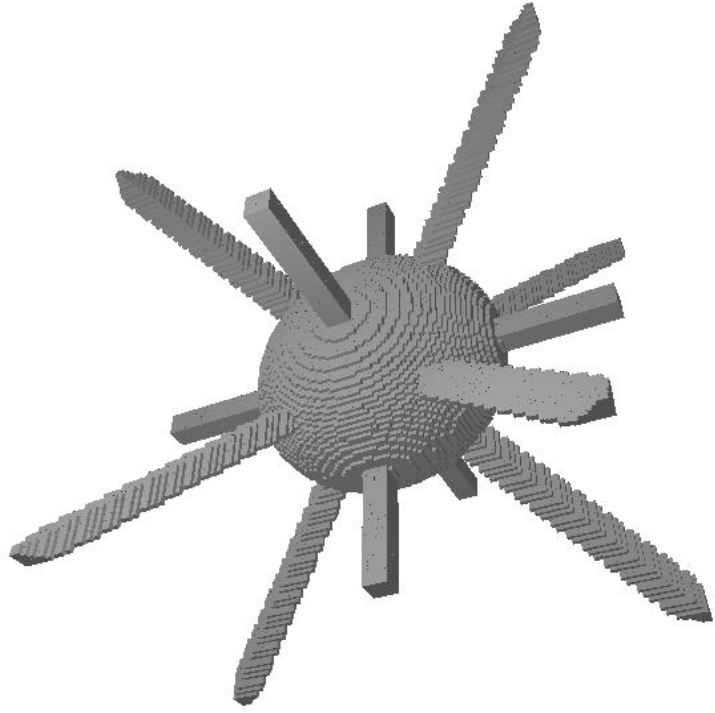

Fig. 1. The optical model of a platelet. This is an oblate spheroid with 14 rectilinear parallelepipeds of square cross-section coming from the centre. The figure corresponds to the ratio of pseudopodia to spheroid volume of 19.6%.

## 5. T-matrix simulation

We applied the T-matrix method to simulate light scattering of a platelet modeled as an oblate spheroid. A recent review of the T-matrix approach can be found in Ref. [16]. We applied the public-domain T-matrix code from Mishchenko (http://www.giss.nasa.gov/crmim/).

## 6. Experimental Equipment and Procedures

The experimental part of this study was carried out by means of the Scanning Flow Cytometer (SFC) that allows measurement of the angular dependency of light-scattering intensity in the region ranging from $5^0$ to $120^0$. The current set-up of the SFC provides measurement of the following combination of Mueller matrix elements [7]:

$$I_s(\theta) = \int_0^{2\pi} [S_{11}(\theta,\varphi) + S_{14}(\theta,\varphi)]d\varphi, \qquad (1)$$

where $I_s(\theta)$ is the output signal of the SFC, $\theta$ and $\varphi$ are the polar and azimuthal angles, respectively. The integration over azimuthal angle $\varphi$ of the second term in Eq. (1) is zero [9]. Therefore, the SFC output signal is proportional to $S_{11}$ integrated over azimuthal angle.

We continuously measured 10.000 LSPs of platelets with the SFC. Each of them was compared with a large set of theoretical LSPs (over a range of parameters) by a $\chi^2$-test:

$$\chi^2 = \frac{\sum_{i=1}^{N}(\log(I_{\exp}(\theta_i)+1) - \log(I_{theor}(\theta_i)+1))^2}{N}, \qquad (2)$$

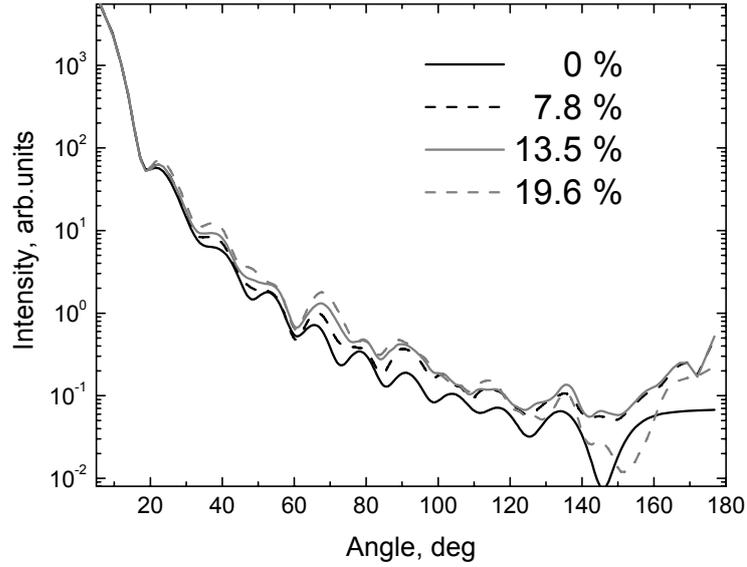

Fig. 2. The LSPs of an ordinary spheroid and spicular spheroids with 14 pseudopodia of different thickness. Calculation have been done for a spheroid with a volume of 6.87 $\mu m^3$, an aspect ratio of 1.4, and a refractive index of 1.39, corresponding to a size parameter equal to 15.0 and phase shift equal to 1.28 assuming a wave length of 0.66 $\mu m$. Due to different thicknesses of the pseudopodia the ratio of pseudopodia to spheroid volume was different: 0.0%, 7.8%, 13.5%, 19.6%.

where $I_{exp}(\theta_i)$ is the experimental LSP measured at $N$ angular values $\theta_i$ ($\theta_1 = 10^0$, $\theta_N = 45^0$), $I_{theor}(\theta_i)$ is the theoretical LSP calculated at the same angles. The comparison was only up to 45 degrees because of the low signal-to-noise ratio for the light-scattering signal above 45 degrees. The theoretical curve that gave the smallest $\chi^2$ value was assumed to be the 'best fit' of the experiment, and the volume, aspect ratio and refractive index for this 'best fit' where assigned to the measured platelet.

## 7. Results and discussions

We have studied the sensitivity of the LSP to variations in volume of the pseudopodia, while keeping the volume of the spheroid fixed. The DDA results are shown in Fig. 2. Calculations have been performed for a spheroid with a volume of 6.87 $\mu m^3$, an aspect ratio of 1.4, a refractive index of 1.39, corresponding to a size parameter $\alpha$ and phase-shift parameter $\rho$ of 15.0 and 1.28, respectively, where $\rho=2\alpha(m-1)$, $\alpha=2\pi r/\lambda \cdot n_0$, $m=n/n_0$ is the particle relative refractive index, $n$ is the refractive index of the particle, $n_0$ is the refractive index of medium, $r$ is the radius of volume equivalent sphere, $\lambda$ is the wavelength. Due to different thickness of the pseudopodia the ratio of pseudopodia and spheroid volumes was different. A ratio of 0.0 % corresponds to absence of pseudopodia, 7.8 % corresponds to pseudopodia thickness of 0.25 $\mu m$, 13.5 % corresponds to a pseudopodia thickness of 0.31 $\mu m$, and 19.6 % corresponds to a pseudopodia thickness of 0.33 $\mu m$. We found small differences in the structure of LSPs and more evident differences in amplitude, increasing with angle in our measuring range of 10 to 45 degrees. For side and back scattering the structure and amplitude of LSPs significantly depend on the presence and volume of the pseudopodia. Based on these results, we conclude that an oblate spheroid can be used to model the intensity of light scattering of platelets for scattering angles smaller then 60 degrees

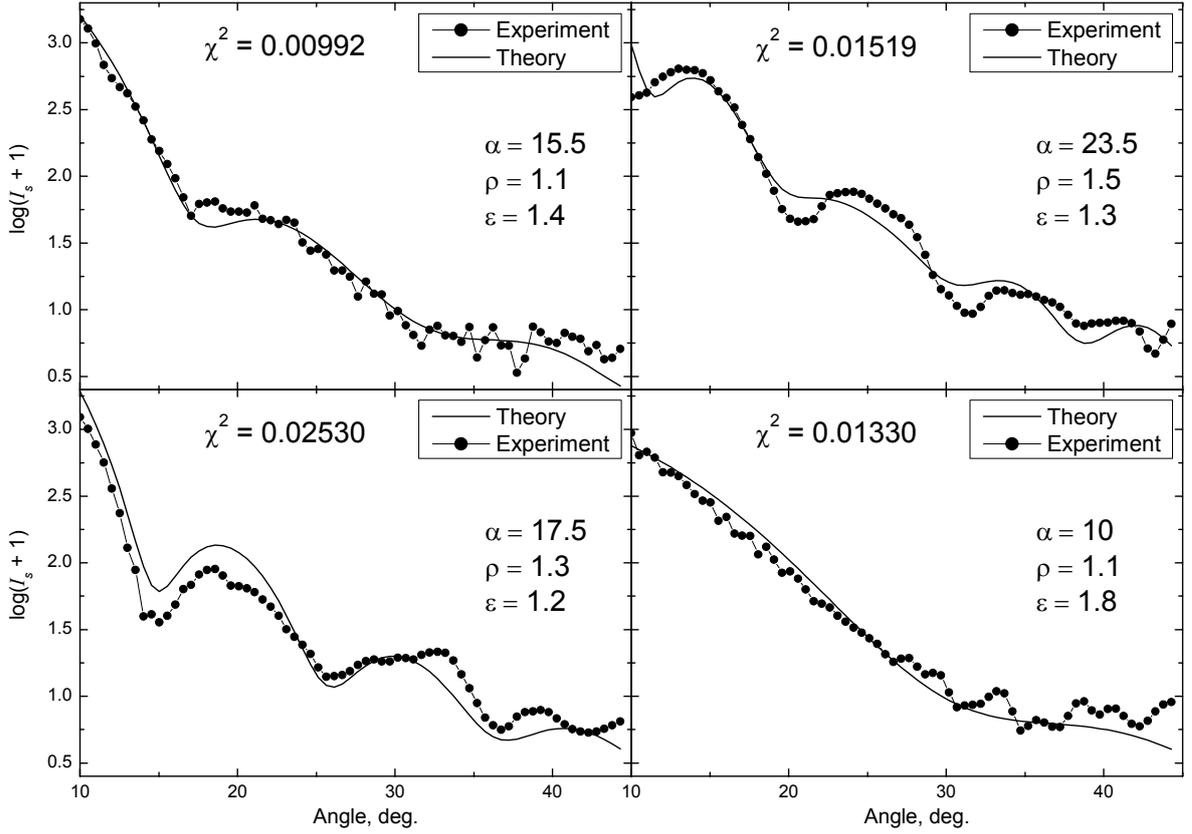

Fig. 3. Experimental and theoretical LSPs of individual platelets. Values of $\chi^2$ differences are shown on the plots.

To determine the characteristics of blood platelets the inverse light-scattering problem must be solved. A direct fitting of the experimental traces with T-matrix computations cannot be used because a T-matrix calculation requires approximately 1 minute for a single set of fixed parameters. We apply the knotted-space approach, where a large set of LSPs, over a range of different volumes, aspect ratios and refractive indices were pre-computed, and compared against experimental traces, as explained in section 6.

The features of the hydrodynamic system of the SFC provide primary orientation of a non-spherical particle with its long axis along flow [17, 18]. Moreover due to small size parameter of platelets and the relatively small angular interval for the measuring angles the orientation of the oblate spheroid does not change the LSP substantially. We verified this with T matrix computations for a few spheroid parameters (data not shown). Therefore, we used this one single orientation in our computations.

A refractive index $m$ relates to fundamental optical characteristic of a particle. However we showed earlier [19] that the inverse light-scattering problem can be effectively solved for the phase-shift parameter. The range of parameters that we used in the computations was for the size parameter $\alpha$ from 7 to 26 with a step of 0.5, for the phase-shift parameter $\rho$ from 0.7 to 2.6 with a step of 0.1, and for the aspect ratios $\varepsilon$ from 1 to 2.4 with a step of 0.1. These values correspond to a volume ranging from 0.7 $\mu m^3$ to 36 $\mu m^3$ and a refractive index ranging from 1.36 to 1.44 (with a wavelength of 0.66 $\mu m$ and refractive index of the medium of 1.333).

A representative result of comparison of experimental and theoretical LSPs by the $\chi^2$-test is shown in Fig. 3. The theoretical LSP fits the experimentally measured LSP quite good. In order to determine the characteristic of measured platelets from light scattering we used the knotted-space and $\chi^2$-test. The LSPs of thousand platelets were processed with this approach and the resulting distributions of platelets in volume, refractive index, and aspect ratio are shown in Fig. 4, Fig. 5, and Fig. 6. The knotted space method with current steps in oblate spheroid characteristics provides a mean $\chi^2$ of 0.026. We are able to state that profile of the volume distribution looks like lognormal distribution that is typical for platelets volume. The parameters of the volume distribution were compared with results obtained from the commercial cell counter Coulter MAX M. The patient platelets were characterized by a mean platelet volume (MPV) of 7.8 fl and 9.3±0.6 fl from MAX M and SFC, respectively. Measured volume distributions allow evaluation of additional parameter, platelet distribution width (PDW). MAX M and SFC gave PDW of 15.4 and 17.4±3, respectively. MPV is considered to define type of vascular diseases; aspect ratio is an important parameter for examination of process of platelet activation; refractive index of a platelet can be referred to concentrations of any substance in platelet.

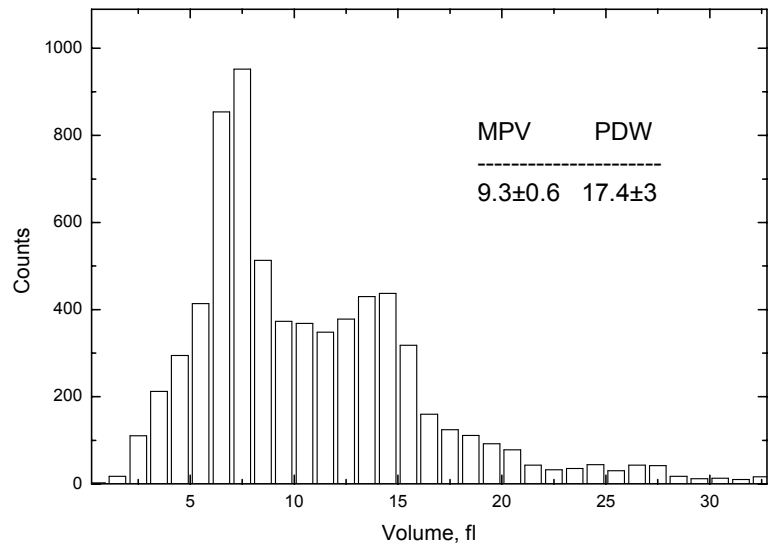

Fig. 4. Distribution of platelets over volume obtained using $\chi^2$-test.

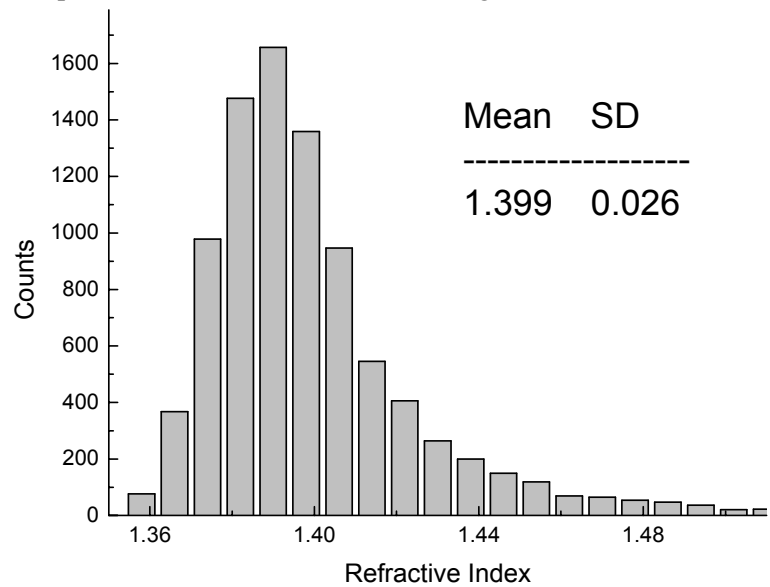

Fig. 5. Distribution of platelets over refractive index obtained using $\chi^2$-test.

Additionally the LSPs for spicular spheroids calculated from DDA were processed with the $\chi^2$-test with the LSPs for smooth oblate spheroids. The results presented in Table 1 demonstrate an effect of pseudopodia on the $\chi^2$ value. There is the weak correlation between the ratio of pseudopodia/spheroid volume and phase-shift parameter. From this we can conclude that the refractive index of the blood platelets has been measured with a small systematical error. This error does not exceed 0.005 for refractive index calculation. On the other hand the $\chi^2$ for LSP with a ratio of pseudopodia/spheroid volume of 19.6 % (Table 1) is three times less then the mean $\chi^2$ for all blood platelets measured and processed with $\chi^2$-test. We assume that experimental noise, deviation from the perfect spheroid shape, and inhomogeneity of blood platelets are a more significant error sources.

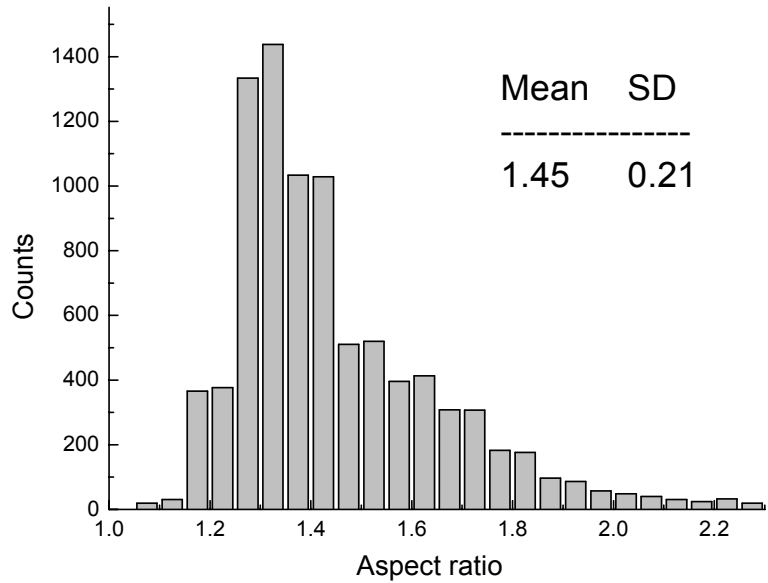

Fig. 6. Distribution of platelets over aspect ratio obtained using $\chi^2$-test.

Table 1. Parameters and values of $\chi^2$ for LSPs with different pseudopodium volume obtained from $\chi^2$-test. The characteristics of spicular spheroid are as follows: size parameter of 15.0, phase shift of 1.28, aspect ratio of 1.4. The numbers in the fist column are the ratio of pseudopodia volumes and spheroid volume.

|   | α | ρ | ε | $\chi^2$ |
|---|---|---|---|---|
| 0 % | **15** | **1.3** | **1.4** | 0.0002 |
| 7.8 % | **14.5** | **1.4** | **1.3** | 0.0018 |
| 13.5 % | **15** | **1.4** | **1.4** | 0.0028 |
| 19.6 % | 14.5 | 1.5 | 1.3 | 0.0065 |

## 8. Conclusion

We introduce a new method to determine volume, shape and refractive index of individual blood platelets. The method is based on Scanning Flow Cytometry with measurement of multi-angle light scattering and solution of the inverse light-scattering problem with the knotted-space approach. We have considered two optical models (spicular and smooth spheroids) of blood platelets. Simulations with DDA have shown that pseudopodia of the platelets activated by EDTA modified the light-scattering pattern substantially for scattering angles above 60 degrees. This fact was in agreement with the angular operational range of the SFC adjusted for measurement of light-scattering patterns of individual blood platelets. We have applied the knotted-space approach to retrieve platelet characteristics from measured light-scattering patterns in real-time. The parameters of platelet volume distribution are in good agreement with independent measurements from commercial instrument and literature values.

This work gives access to new hematological indices which can be used as a diagnostic indicator of pathologies. A clinical protocol should be developed for the scanning flow cytometer with the aim of measuring the mean and the width of distribution of blood platelet density (refractive index), the mean and the width of distribution of platelet asphericity.

**Acknowledgment**

This research was supported by Russian Foundation for Basic Research through the grant 03-04-48852-a, by Siberian Branch of the Russian Academy of Sciences through the grant 115-2003-03-06, and by the NATO Science for Peace program through grant SfP 977976.